\documentclass[sigconf]{acmart}
\AtBeginDocument{%
  }

\copyrightyear{2025}
\acmYear{2025}
\setcopyright{cc}
\setcctype{by-sa}
\acmConference[CIKM '25]{Proceedings of the 34th ACM International
Conference on Information and Knowledge Management}{November 10--14,
2025}{Seoul, Republic of Korea}
\acmBooktitle{Proceedings of the 34th ACM International Conference on
Information and Knowledge Management (CIKM '25), November 10--14, 2025,
Seoul, Republic of Korea}\acmDOI{10.1145/3746252.3761405}
\acmISBN{979-8-4007-2040-6/2025/11}

\begin{document}

\title{As Good as It KAN Get: High-Fidelity Audio Representation}

\author{Patryk Marszałek}
 \authornote{Both authors contributed equally to this research.}
\orcid{0009-0008-7695-7905}
 \affiliation{%
   \institution{Jagiellonian University}
   \city{Krakow}
   \country{Poland}
 }
  \email{patryk.marszalek@student.uj.edu.pl}

\author{Maciej Rut}
\authornotemark[1]
\orcid{0009-0009-9637-6128}
 \affiliation{%
   \institution{Jagiellonian University}
   \city{Krakow}
   \country{Poland}
 }
 \email{maciej.rut@student.uj.edu.pl}

\author{Piotr Kawa}
\orcid{0000-0002-2025-0547}
 \affiliation{%
   \institution{Wroclaw University of Science and Technology}
   \city{Wroclaw}
   \country{Poland}
 }
 \email{piotr.kawa@pwr.edu.pl}

\author{Przemysław Spurek}
\orcid{0000-0003-0097-5521}
\affiliation{%
   \institution{Jagiellonian University}
   \city{Krakow}
   \country{Poland}
}   
\affiliation{%
   \institution{IDEAS Research Institute}
   \city{Warszawa}
   \country{Poland}
 }
 \email{przemyslaw.spurek@uj.edu.pl}

 \author{Piotr Syga}
 \orcid{0000-0002-0266-5802}
 \affiliation{%
   \institution{Wroclaw University of Science and Technology}
   \city{Wroclaw}
   \country{Poland}
 }
 \email{piotr.syga@pwr.edu.pl}

\renewcommand{\shortauthors}{Patryk Marszałek, Maciej Rut, Piotr Kawa, Przemysław Spurek, and Piotr Syga}

\begin{abstract}
Implicit neural representations (INR) have gained prominence for efficiently encoding multimedia data, yet their applications in audio signals remain limited. This study introduces the Kolmogorov-Arnold Network (KAN), a novel architecture using learnable activation functions, as an effective INR model for audio representation. KAN demonstrates superior perceptual performance over previous INRs, achieving the lowest Log-Spectral Distance of 1.29 and the highest Perceptual Evaluation of Speech Quality of 3.57 for 1.5~s audio. To extend KAN's utility, we propose FewSound, a hypernetwork-based architecture that enhances INR parameter updates. FewSound outperforms the state-of-the-art HyperSound, with a 33.3\% improvement in MSE and 60.87\% in SI-SNR. These results show KAN as a robust and adaptable audio representation with the potential for scalability and integration into various hypernetwork frameworks.
\end{abstract}






\begin{teaserfigure}
  \centering
    \includegraphics[width=0.85\textwidth]{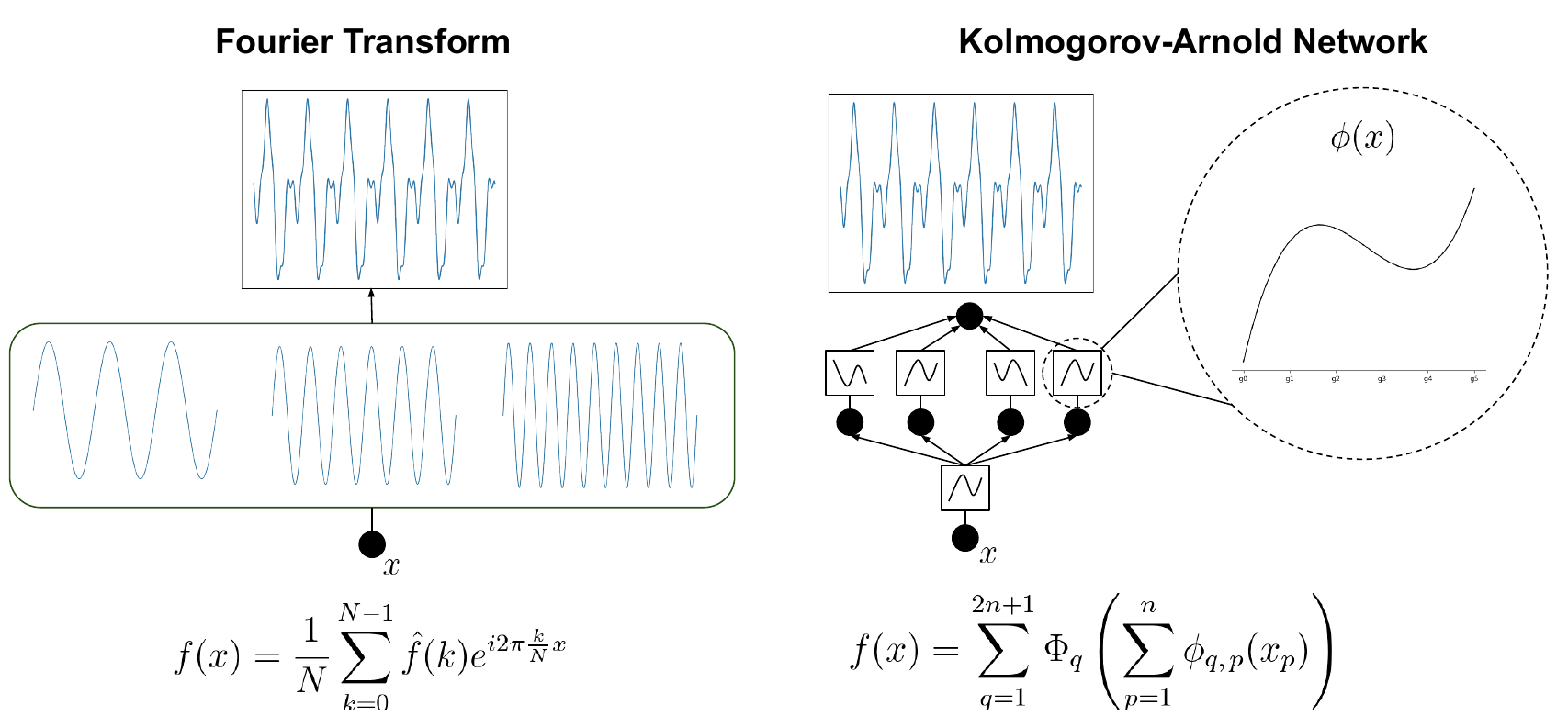}
    \Description{Sound representation is a challenging task. In classical analysis, we usually apply the FFT (Fast Fourier Transform), which analyzes frequencies by separating them into sinusoidal components. We propose to analyze sounds using KAN (Kolmogorov-Arnold Network), which uses trainable activation functions instead of classical layers. Such an operation can be seen as approximating signals by a linear combination of basis functions, bearing a similarity to the FFT or wavelets.}
    \caption{Sound representation is a challenging task. In classical analysis, we usually apply the FFT (Fast Fourier Transform), which analyzes frequencies by separating them into sinusoidal components. We propose to analyze sounds using KAN (Kolmogorov-Arnold Network), which uses trainable activation functions instead of classical layers. Such an operation can be seen as approximating signals by a linear combination of basis functions, bearing a similarity to the FFT or wavelets.}
    \label{fig:trajectorid}
\end{teaserfigure}




\def\Loss{\mathcal{L}}
\def\R{\mathbb{R}}
\def\our{FewSound}


\maketitle

\section{Introduction} 

Implicit neural representations (INRs) depict discrete data in neural network weights. INR finds many applications in the case of image~\cite{klocek2019hypernetwork}, video~\cite{video-super-resolution}, or 3D objects~\cite{spurek2020hypernetwork}. In the case of sound, INRs are unexplored, and only a few approaches are dedicated to such data. HyperSound~\cite{szatkowski2023hypernetworks} presents an autoencoder-based architecture that takes raw input sound, and produces weight for INR-based sound representation. INRAS~\cite{inras} is an INR network that, by mapping the scene coordinates to the corresponding impulse responses, enables the modeling of spatial audio of interior scenes. Siamese SIREN \cite{siamese_siren} is the extension of SIREN \cite{SIREN} used for audio compression.

The recently proposed KAN (Kolmogorov-Arnold Network) \cite{liu2024kan} replaces traditional fully connected layers with learnable activation functions based on splines. Any function f (e.g., an audio representation of the magnitude over time) can be approximated by splines with $\sup\|\text{f-spline}\|=\mathrm{O}(h^4)$~\cite{hall1976optimal}. KANs inherit this theoretical precision as a linear combination of splines, resulting in a polynomial bound. This indicates potential effectiveness for audio reconstruction.
While KAN shows potential, it faces limitations with high-dimensional data. However, its effectiveness in low-dimensional settings and successful application to complex function regression tasks make it a promising candidate for audio INR.

In our work we examine six INR approaches: SIREN \cite{SIREN}, NeRF \cite{mildenhall2020nerf}, Random Fourier Features \cite{FourierFeatures}, WIRE~\cite{WIRE}, FINER \cite{liu2024finer} and KAN \cite{liu2024kan}.
This paper shows that KAN is effective in modeling individual audio signals and can be extended to various functions through the hypernetwork paradigm. For single audio INR, KAN trains more rapidly and achieves better results.
We also show that KANs can be generated by hypernetworks. 
In practice, we can directly produce KAN parameters or use updates for a universal KAN-based architecture.
Hence, we propose two methods. The first one is classical HyperSound~\cite{szatkowski2023hypernetworks} with a KAN target network, while the second one uses a few-shot setting \cite{tancik2021learned,batorski2024hyperplanes} in INR. As there is no few-shot model for audio signals, we introduce \our{}.
\footnote{The source code for the proposed method can be found at the following link: \href{https://github.com/gmum/fewsound}{https://github.com/gmum/fewsound}}
It takes raw sound on input and uses a hypernetwork to generate updates to the universal weights of an INR dedicated to encoding sound.
\our{} is an implementation of the Model-Agnostic Meta-Learning algorithm (MAML) \cite{finn2017model}.
Thus, two settings show that KAN can be effectively modeled by hypernetworks. 


Key contributions of our research include:
\begin{itemize}
    \item introducing KAN-based sound representation and showing its efficacy,
    \item introducing \our{}, a hypernetwork-based meta-learning method, which is used with KAN networks and improves over previous hypernetwork approach HyperSound~\cite{szatkowski2023hypernetworks},
    \item comparing KAN represenation with other INRs, 
    \item showing that KAN can be produced via hypernetworks,
    \item analysis of the parameters' influence on the efficacy of the proposed method.
\end{itemize}

\section{Related Work}\label{sec:related}

To our knowledge, there are no specialized INR architectures focused on sound. Research papers on INR \cite{SIREN,FourierFeatures,WIRE,liu2024finer,instantNGP} typically evaluate models using images, videos, and 3D objects as benchmarks. Although experiments involving sounds do exist, they remain relatively unexplored. In this section, we show that KAN effectively models single-audio in an efficient manner.

\paragraph{\textbf{INR}}
Implicit neural representations (INR) are data approximations based on neural networks relying on coordinate inputs. They enable discrete information such as pixels~\cite{klocek2019hypernetwork,video-super-resolution}, cloud points~\cite{spurek2020hypernetwork}, or audio magnitude information~\cite{szatkowski2023hypernetworks} to be represented as continuous functions. The separation of the representation from spatial resolution makes the representation more memory efficient. Such information is limited by the networks' architecture instead of the grid resolution. INRs are used in many signal processing problems, including 3D reconstruction~\cite{mildenhall2020nerf}, super-resolution~\cite{video-super-resolution}, compression~\cite{inr-compression,siamese_siren} or audio scene representation~\cite{inras}. 

Initially, the INRs used ReLU-based MLPs. However, due to the ReLU's second derivative being zero, they cannot capture the fine details of the information located in higher-order derivatives leading to worse performance. To counteract this, methods as SIREN~\cite{SIREN} have been developed, which use a periodic activation function in place of ReLU. 
The i-th layer of the network is defined as: $\phi_{i}(x_{i}) = \sin(\omega_{i} W_{i}^Tx_{i}+b_{i}),$
where $x_i$ is layer's input, while $W_i$ and $b_i$ are, respectively, weights and biases.
The use of the sine function enables SIREN to capture the details of the processed signals to a much greater degree than the widely used activation functions.
Another example of a network architecture using a specialized activation function is WIRE (wavelet implicit neural representation) \cite{WIRE}. Inspired by harmonic analysis, WIRE uses complex Gabor wavelet $\psi$ as its nonlinearity activation:
$\psi(x) = e^{i\omega_0x}e^{-\lvert s_0x \rvert^2}$.

INRs are prone to \textit{spectral bias}~\cite{spectral-bias}, where MLPs favor low-frequency functions and neglect high-frequency signals, reducing performance when such components are informative~\cite{gorji2023}. This is commonly mitigated with high-frequency embeddings~\cite{mildenhall2020nerf, FourierFeatures}, which reshape the input space and adjust output frequencies.

FINER~\cite{liu2024finer}, an MLP-based architecture, flexibly tunes spectral bias via a variable-periodic activation $a(x) = \sin(\omega_0 \lvert x + 1 \rvert x)$ and specialized first-layer bias initialization.

NeRF (Neural Radiance Field)~\cite{mildenhall2020nerf} represents scenes with a 5D input: spatial location $(x,\,y,\,z)$ and viewing direction $(\theta,\,\phi)$, and an MLP that outputs color $c = (r,\,g,\,b)$ and density $\sigma$, enabling 3D reconstruction from sparse views.

\paragraph{\textbf{Sound representation}}
In our research, we use learned representations from deep neural networks, which outperform traditional hand-crafted features~\cite{baevski2020wav2vec}. These representations fall into 2 categories: acoustic embeddings from neural speech codecs~\cite{encodec}, and semantic embeddings~\cite{hubert}. We focus on representations derived from models designed for ASR or related tasks.

SoundStream \cite{soundstream} is a popular neural audio codec based on the encoder-decoder architecture. In our experiments we used only the encoder part composed of 1D convolution followed by four encoder blocks with residual units and the final 1D convolution.

SNAC~\cite{snac} employs a hierarchical token-based encoder, similar to prior work, but samples coarse tokens less frequently to span longer time windows. This reduces reconstruction bitrate and aids language modeling tasks.

Whisper \cite{whisper} is a Transformer-based \cite{transformer} ASR system, trained on 680,000 hours of data, achieving state-of-the-art speech recognition results. It was successfully adapted to various other tasks including audio-language models \cite{qwen}, voice conversion \cite{whisper-vc}, text-to-speech \cite{whisper-tts} and audio DeepFake detection \cite{deepfake-whisper-features}. In our experiments, we use only Whisper's encoder.

Spectrogram Encoder is based on VoiceBlock \cite{voiceblock} --- originally developed as a speech-to-speech solution for adversarial examples against ASR models. The encoder is based on the linear-spectrogram input that extracts a feature vector composed of fundamental frequency, aperiodicity and loudness of the signal.

\begin{figure*}[!h]
    \centering
    \includegraphics[width=0.9\textwidth]{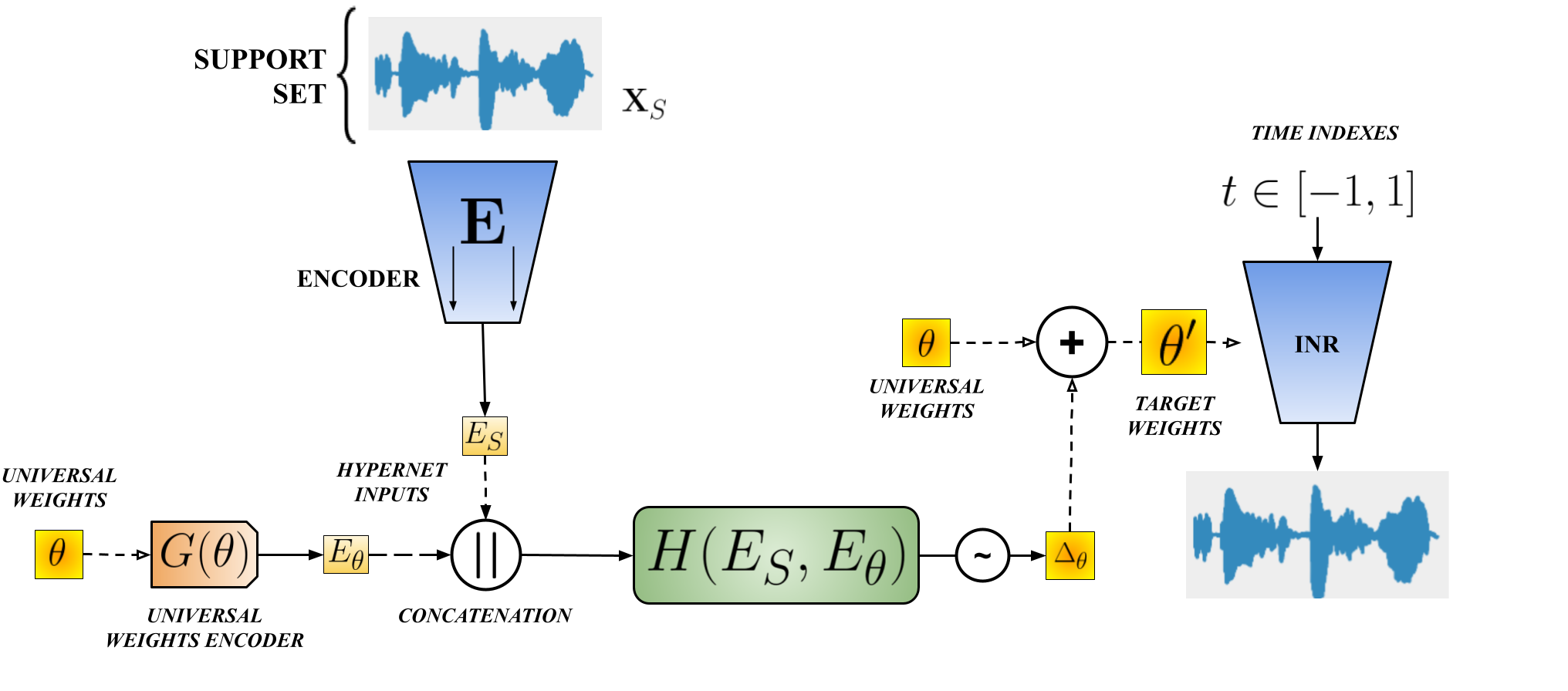}
    \Description{The overview of \our{} architecture. The input raw sound is processed by encoding network $E(\cdot)$ and, together with encoded universal weights, passed to the fully connected Hypernetwork $H(\cdot)$. The Hypernetwork transforms them and returns the update of weights $\Delta \theta$ for the target INR $f_{\theta'}$. Finally, the target network $f_{\theta'}$ maps time samples to sound amplitudes specific to the considered task.}
    \caption{
    The overview of \our{} architecture. The input raw sound is processed by encoding network $E(\cdot)$ and, together with encoded universal weights, passed to the fully connected Hypernetwork $H(\cdot)$. The Hypernetwork transforms them and returns the update of weights $\Delta \theta$ for the target INR $f_{\theta'}$. Finally, the target network $f_{\theta'}$ maps time samples to sound amplitudes specific to the considered task.  
    }
    \label{fig:architecture}
\end{figure*}

\paragraph{\textbf{KAN (Kolmogorov-Arnold Network)}}
KAN~\cite{liu2024kan} is a type of neural network architecture used as an alternative to MLPs, instead of using fixed activation functions. KANs utilize learnable activation functions placed on edges rather than nodes. KANs' learnable functions are parametrized as a spline and serve as ''weights'', whereas the nodes only sum the incoming signals. This is described by the following formula: 
\begin{displaymath}
f(x) =\sum_{q=1}^{2n+1} \Phi_q\left(\sum_{p=1}^n\phi_{q,\,p}(x_p)\right)~,
\end{displaymath}
where $\Phi_{q}$ is a KAN layer, $\phi_{q,\,p}$ is a single activation function and $x_p$ is input feature. The authors introduced LANs (learnable activation networks), which differ from MLPs only by learning activation functions. They demonstrate that such networks can be applied to implicit neural representation tasks.

\paragraph{\textbf{Audio INR}}
\label{par:audio_inr}
In the realm of acoustic data, INR transforms a one-dimensional temporal space into the corresponding amplitude values. Therefore, our INR is depicted as:
\begin{displaymath}
f_{\theta}: \R \to \R~,
\end{displaymath}
where \( f_{\theta}(t) \) denotes the amplitude of the sound over time.

We consider six variants of INRs: NeRF, SIREN, RFF, FINER, WIRE and KAN.
We use NeRF to refer to a network with positional encoding, followed by an MLP with ReLU activation. The positional encoding, described by $$\gamma(t) = \left[ \sin(2^0\pi t), \cos(2^0\pi t), \ldots,
        \sin(2^{L-1}\pi t), \cos(2^{L-1}\pi t) \right]$$
is taken from the original NeRF \cite{mildenhall2020nerf}.

SIREN \cite{SIREN} is an MLP with sine activation: $a(x) = \sin(\omega_0x)$, where $\omega_0$ is a hyperparameter.

RFF (Random Fourier Features) \cite{FourierFeatures} is used to refer to an MLP with ReLU activation and input embedding defined as $$\gamma(t) = \left[ \cos(2 \pi \mathbf{B}t),
    \sin(2 \pi \mathbf{B} t)\right],$$
where $\mathbf{B} \in \mathbb{R}^{m \times d}$ is a random projection matrix, sampled from $\mathcal{N}(0, \sigma^2)$. In our case $d$, the dimensionality of the input, is always equal to 1, while $\sigma$ and $m$ are hyperparameters.

WIRE \cite{WIRE} is an MLP with complex Gabor wavelet activation $a(x) = e^{i\omega_0x}e^{-\lvert s_0x \rvert^2}$, where $\omega_0$ and $s_0$ are hyperparameters.

FINER \cite{liu2024finer} is an MLP with activation function defined as $a(x) = \sin(\omega_0 \lvert x + 1 \rvert x )$, where $\omega_0$ is a hyperparameter.
The bias of its first layer is initialized by sampling from $\mathcal{U}(-k, k)$, where $k$ is a hyperparameter.

Our Kolmogorov-Arnold network implementation consists of the same positional encoding as in NeRF, followed by KAN layers. Each KAN layer can be described by $\phi(x) = w_b\mathrm{SiLU}(x) + w_sspline(x),$
where $w_b$ and $w_s$ are matrices of learnable weights, and $spline(x)$ is a linear
combination of B-splines such that $spline(x) = \sum_i{c_iB_i(x)}$
with learnable $c_i$s. It is parametrized by \textit{spline order} -- the degree of the spline, and \textit{grid size} -- the number of knots defining the spline.

\section{KAN in hypernetwork-based architectures}

In this section, we demonstrate how KAN integrates with hypernetworks using two main approaches. First, we extend the autoencoder-based HyperSound~\cite{szatkowski2023hypernetworks} by employing KAN as the target model. Second, we explore KAN in a few-shot setting introduced by~\cite{wang2020generalizing}. To this end, we propose \our{}, which adapts the framework of \cite{wang2020generalizing} to the context of KANs and hypernetworks.


\paragraph{\textbf{Hypernetwork}}

In the seminal work~\cite{ha2016hypernetworks}, Hypernetworks are introduced as neural models that produce the weights of another network dedicated to solving a specific task. The approach aims to reduce the number of trainable parameters by employing a hypernetwork with fewer parameters than the target network. Building on this concept, numerous studies have emerged. Among them, \cite{sheikh2017stochastic} draws an analogy to generative models, demonstrating how this technique can produce a variety of target networks approximating the same function. In the audio INR domain, HyperSound~\cite{szatkowski2023hypernetworks} leverages hypernetworks to transform raw audio into weights of an INR-based representation.


\paragraph{\textbf{Few-shot learning for INR}}

A key limitation of deep learning models is their dependence on large datasets, whereas humans can learn effectively from just a few examples. Few-shot learning~\cite{wang2020generalizing} seeks to bridge this gap by structuring data into numerous small tasks and using architectures capable of rapid adaptation to new problems in a single step. The most popular method, MAML~\cite{finn2017model}, leverages a set of universal weights that are fine-tuned for specific small tasks through a limited number of gradient updates. Its model-agnostic nature allows it to be applied to a wide range of models and learning strategies. Nevertheless, further generalization is possible. HyperMAML~\cite{PRZEWIEZLIKOWSKI2024128179} achieves this by replacing the gradient-based adaptation step with a Hypernetwork that predicts task-specific weight updates directly from the data.


The terminology of few-shot learning is often inconsistent due to differing definitions in prior work. Standardized frameworks exist~\cite{chen2019closer,wang2020generalizing}, but their nomenclature is focused on a classification, regression, or generative task. For INR-based models, we follow~\cite{tancik2021learned}. In this setting, we have a dataset $X$ of sound samples and aim to produce an INR $f_{\theta}$ that encodes the audio signals in its weights.


\paragraph{\textbf{\our{}}}

At this stage, we are ready to present \our{}, a model that employs Hypernetworks to adapt universal INR weights specialized for sound. As illustrated in Fig.~\ref{fig:architecture}, its key idea lies in integrating the information extracted from raw sound data and universal weights to determine the optimal adjustment for a particular task. As a result, our method is able to generate INR's parameters for entirely different sounds.


Within the \our{} framework, we distinguish a trainable encoding network $E(\cdot)$ that maps raw audio data to a low-dimensional representation $\mathbf{E}_{\mathcal{S}}$. We explore several encoder architectures, including SoundStream, Whisper, SNAC, and Spectrogram Encoder, and report their performance in Tab.~\ref{table:encoder-comparison}. In addition, we define a universal weight encoder~$\mathbf{G}(\cdot)$ that compresses universal weights~$\theta$ to a representation~$\mathbf{E}_{\mathcal{\theta}}$ of the same dimensionality as $\mathbf{E}_{\mathcal{S}}$. Both representations are concatenated and subsequently processed by a fully connected hypernetwork $H(\cdot)$, which outputs the adjustment~$\Delta_{\theta}$ to the universal weights for the given input task. Consequently, the final target model parameters~$\theta'$ can be expressed as: 
\begin{equation}
 \theta' = \theta  + \Delta_{\theta} = \theta + H( \mathbf{E}_{\mathcal{S}},\,\mathbf{E}_{\theta}).    
 \label{eq:hyper_update}
\end{equation}
Using these adapted weights, we can reconstruct the training example by running one of the INRs described in Section~\ref{par:audio_inr} over all time indices of the input signal. Although \our{} accepts a fixed-size input (in our case 32768), it can also be used on audio samples of arbitrary length by reconstructing them piece by piece in an overlapping window scheme and applying smoothing during concatenation.


\paragraph{\textbf{Training}} In the training procedure, we assume that the encoder $E(\cdot)$ is parameterized by $\gamma$ ($E:=E_{\gamma}$), the universal weight encoder $G(\cdot)$ by $\delta$, and the Hypernetwork $H(\cdot)$ by $\eta$ ($H:=H_{\eta}$). During each training iteration, we first sample the batch of sounds $\mathcal{B}$ from the dataset $X$. Next, the hypernetwork computes the task-specific update $\Delta_{\theta_i}$, yielding updated weights $\theta'_i$ as given in~\eqref{eq:hyper_update}. These parameters define the corresponding INR $f_{\theta'_i}$. Finally, we obtain the reconstruction~$\hat{x}$ of the input sound~$x$ using time steps from $[-1,,1]$, and train the system parameters with the loss function:

\begin{equation}
\Loss_{\our{}} (f_{\theta}) = \sum_{\mathcal{T}_i \in \mathcal{B}} \Loss(f_{\theta'_i}) = \sum_{\mathcal{T}_i \in \mathcal{B}} \Loss(f_{\theta  + \Delta_{\theta_i}})~,
\label{eq:hyper_global_loss}
\end{equation}

\noindent where $\Loss(f_{\theta'})=\lambda_{t}\Loss_{t}(x,\hat{x}) + \lambda_{f}\Loss_{f}(x, \hat{x})$, with $\Loss_{t}$ denoting L1 loss in temporal space, $\Loss_{f}$ representing multi-resolution mel-scale STFT loss in frequency space~\cite{parallel-wavegan} and the coefficients $\lambda$s controlling the influence of particular losses. The previously defined parameters $\gamma$, $\delta$, and $\eta$, together with the global weights $\theta$, constitute the meta parameters of the system and are updated using AdamW~\cite{adamw} optimizer and one cycle~\cite{one_cycle} learning rate scheduler. 

\begin{table*}
    \centering
    \caption{Reconstruction metrics for 1.5~s and 3~s audio. We report mean and standard deviation across 110 samples, each of a different speaker of VCTK dataset~\cite{vctk}.}
    \label{table:inr-comparison}
    \begin{tabular}{lccccccc}
        \toprule
        & \multicolumn{7}{c}{1.5~s audio}\\
        \midrule
        \textbf{ INR (\# params)} & MSE & PSNR & LSD & SI-SNR & PESQ & STOI & CDPAM \\
        & $\rightarrow 0$ & $\rightarrow \infty$ & $\rightarrow 0$ & $\rightarrow 100$ & $\rightarrow 4.5$ & $\rightarrow 1$ & $\rightarrow 0$ \\
        \midrule
        \textit{NERF (33,261)}
        & 8.1e-4 $\pm$ 1.3e-3
        & 39.68 $\pm$ 5.67
        & 1.36 $\pm$ 0.18
        & 17.83 $\pm$ 6.10
        & 3.21 $\pm$ 0.48
        & 0.98 $\pm$ 0.04
        & 0.25 $\pm$ 0.11
        \\
        \textit{SIREN (31,191)} 
        & 4.4e-4 $\pm$ 4.9e-4 
        & 40.82 $\pm$ 3.68 
        & 1.40 $\pm$ 0.26 
        & 19.71 $\pm$ 2.67
        & 2.76 $\pm$ 0.43
        & 0.95 $\pm$ 0.06
        & 0.23 $\pm$ 0.14
        \\
        \textit{FINER (31,191)} 
        & 1.9e-4 $\pm$ 2.3e-4
        & 44.39 $\pm$ 3.61
        & 2.06 $\pm$ 0.27
        & 23.29 $\pm$ 4.29 
        & 2.26 $\pm$ 0.29
        & 0.93 $\pm$ 0.08
        & 0.51 $\pm$ 0.11
        \\
        \textit{WIRE (31,191)} 
        & 4.8e-4 $\pm$ 6.5e-4
        & 41.91 $\pm$ 5.53
        & 1.63 $\pm$ 0.25
        & 20.77 $\pm$ 5.91
        & 2.79 $\pm$ 0.44
        & 0.96 $\pm$ 0.05
        & 0.16 $\pm$ 0.08
        \\
        \textit{RFF (39,201)} 
        & 8.7e-4 $\pm$ 8.8e-4
        & 38.63 $\pm$ 4.93
        & 1.56 $\pm$ 0.16
        & 17.81 $\pm$ 5.58
        & 2.94 $\pm$ 0.39
        & 0.98 $\pm$ 0.01
        & 0.19 $\pm$ 0.09
        \\
        \textit{\textbf{KAN} (33,768)} 
        & 6.7e-4 $\pm$ 2e-3
        & 42.44 $\pm$ 6.72
        & 1.29 $\pm$ 0.22
        & 20.50 $\pm$ 6.94
        & 3.57 $\pm$ 0.42
        & 0.99 $\pm$ 0.02
        & 0.13 $\pm$ 0.08
        \\
        \midrule
        & \multicolumn{7}{c}{3~s audio} \\
        \midrule
        \textbf{ INR (\# params)} & MSE & PSNR & LSD & SI-SNR & PESQ & STOI & CDPAM \\
        & $\rightarrow 0$ & $\rightarrow \infty$ & $\rightarrow 0$ & $\rightarrow 100$ & $\rightarrow 4.5$ & $\rightarrow 1$ & $\rightarrow 0$ \\
        \midrule
        \textit{NERF (33,261)} 
        & 4.6e-4 $\pm$ 5.0e-4
        & 40.83 $\pm$ 4.50
        & 1.47 $\pm$ 0.16
        & 17.54 $\pm$ 4.29
        & 2.54 $\pm$ 0.45
        & 0.94 $\pm$ 0.05
        & 0.30 $\pm$ 0.14
        \\
        \textit{SIREN (31,191)} 
        & 5.3e-4 $\pm$ 3.6e-4 
        & 39.20 $\pm$ 3.35
        & 1.42 $\pm$ 0.18
        & 14.41 $\pm$ 4.44
        & 2.25 $\pm$ 0.28
        & 0.96 $\pm$ 0.03
        & 0.41 $\pm$ 0.14
        \\
        \textit{FINER (31,191)} 
        & 1.9e-4 $\pm$ 1.8e-4
        & 43.77 $\pm$ 3.19
        & 2.21 $\pm$ 0.16
        & 19.05 $\pm$ 3.51
        & 1.89 $\pm$ 0.25
        & 0.95 $\pm$ 0.07
        & 0.45 $\pm$ 0.12
        \\
        \textit{WIRE (31,191)} 
        & 4.3e-4 $\pm$ 3.9e-4
        & 40.30 $\pm$ 3.42
        & 1.52 $\pm$ 0.10
        & 15.45 $\pm$ 4.36
        & 2.33 $\pm$ 0.35
        & 0.95 $\pm$ 0.07
        & 0.33 $\pm$ 0.13
        \\
        \textit{RFF (39,201)} 
        & 5.3e-4 $\pm$ 3.6e-4
        & 39.20 $\pm$ 3.35
        & 1.42 $\pm$ 0.18
        & 14.41 $\pm$ 4.44
        & 2.25 $\pm$ 0.28
        & 0.96 $\pm$ 0.03
        & 0.40 $\pm$ 0.14
        \\
        \textit{\textbf{KAN} (33,768)} 
        & 3.8e-4 $\pm$ 4.4e-4
        & 42.28 $\pm$ 5.31
        & 1.34 $\pm$ 0.19
        & 18.85 $\pm$ 5.08
        & 2.89 $\pm$ 0.51
        & 0.97 $\pm$ 0.03
        & 0.33 $\pm$ 0.15
        \\
        \bottomrule
    \end{tabular}
\end{table*}

\section{Experiments}
This section presents an experimental analysis in two parts: performance comparison of INR architectures and results of our hypernetwork method, \our{}. 

Multiple metrics are used to evaluate the quality of reconstructions. That includes quantitative metrics such as MSE, Log-spectral distance (LSD), PSNR, and SI-SNR. We also use metrics that measure the perceptual quality of the audio signal, such as PESQ \cite{pesq}, STOI \cite{stoi}, and CDPAM \cite{cdpam}. Given that the perception of 'natural' sound is critical to our output, we prioritized the PESQ metric due to its strong correlation with human auditory perception.

\subsection{Implicit Neural Representations}

We compare the aforementioned INR architectures using two subsets of the VCTK \cite{vctk} dataset. Each of the subsets is generated by randomly choosing one recording per speaker, resampling it to 22.05 kHz and randomly cropping it to a desired length. We crop the recordings to 32,768 samples (about 1.5 seconds) in the first subset and to 65,536 (about 3 seconds) in the second subset. Both subsets consist of a total of 110 recordings. To make a fair comparison, all of the MLP based INRs have the same number of layers with the same sizes, while KANs have their architecture chosen in a way that the number of parameters is similar to the other models. All architectures are trained for 10,000 steps, using the AdamW optimizer~\cite{adamw} and the same loss function $\Loss$ as in Eq.~\ref{eq:hyper_global_loss}. Table \ref{table:inr-comparison} shows the results of the experiments. 
One may note that, with the exception of MSE, the shorter audios result in better quality. KAN is competitive for most metrics, achieving the best mean score in LSD, PESQ, STOI and CDPAM and second best PSNR for 1.5~s audio; as well as the best mean LSD, PESQ, STOI, and second best MSE, PSNR, SI-SNR, CDPAM for 3~s audio.

Further comparison of these INRs on additional datasets is available in section~\ref{sec:additional_inr}.

\subsection{Hypernetworks}
\paragraph{\textbf{Setting}}
When using hypernetworks, we only consider NeRF, SIREN, and KAN as the target network architectures.
We compare our method with the state-of-the-art model HyperSound \cite{szatkowski2023hypernetworks} and follow the same experimental setting. We use the VCTK dataset resampled to $22,050$ Hz. The inputs are truncated to length of 32,768 samples. The last 10 speakers are used as the validation set.
All models are trained for 2,500 epochs, with 10,000 training samples chosen randomly for each epoch. We use AdamW optimizer~\cite{adamw} and OneCycleLR scheduler \cite{superconvergence}. The initial learning rate is set to $1\mathrm{e}{-5}$, with the exception of runs using the SIREN target network, where it is $1\mathrm{e}{-6}$.

We examine how different encoder and target network architectures impact the quality of reconstructions. In particular, we explore the performance of KAN-based target networks used both with \our{} and HyperSound models.

Unless otherwise mentioned, \our{} uses a SoundStream~\cite{soundstream} based encoder, 17.5M parameter hypernetwork MLP and NERF-based target network. The number of parameters in target networks is roughly equal to the length of the input audio (32K). Results for HyperSound are taken from the original paper.

\paragraph{\textbf{Results}}

Table \ref{table:sota-comparison} shows the comparison of our model with state-of-the-art HyperSound. We include results for two sizes of the hypernetwork MLP -- a smaller 17.5M parameter model and a larger one with 41.2M parameters. Both of them have comparable results, showing that the smaller network may be used for efficiency. One may note that even the smaller one outperforms HyperSound in all the metrics (MSE by 33.3\%, LSD by 6.83\%, SI-SINR by 60.87\%, PESQ by 8.66\%, STOI by 2.63\% and CDPAM by 13.89\%).

\begin{table}[htbp]
    \centering
    \caption{\our{} performance compared to HyperSound.}
    \label{table:sota-comparison}
    \begin{tabular}{@{}l@{\;\;}c@{\;\;}c@{\;\;}c@{\;\;}c@{\;\;}c@{\;\;}c@{\;\;}c@{}}
        \toprule
        \textbf{Model} & MSE & LSD & SI-SNR & PESQ & STOI & CDPAM \\
        & $\rightarrow 0$ & $\rightarrow 0$ & $\rightarrow 100$ & $\rightarrow 4.5$ & $\rightarrow 1$ & $\rightarrow 0$ \\
        \midrule
        \textit{HyperSound} & 0.009 & 1.61 & 2.99 & 1.27 & 0.76 & 0.36 \\
        \textit{\our} 17.5M & \textbf{0.006} & \textbf{1.50} & 4.81 & 1.38 & 0.78 & 0.31 \\
        \textit{\our} 41.2M & \textbf{0.006} & \textbf{1.50} & \textbf{5.03} & \textbf{1.41} & \textbf{0.79} & \textbf{0.30} \\
        \bottomrule
    \end{tabular}
\end{table}

\paragraph{\textbf{Impact of target network architecture}}
Tables \ref{table:our-target-comparison} and \ref{table:hs-target-comparison} show how different target network architectures impact the quality of reconstructions using \our{} and HyperSound  models, respectively. Note that for \our{} NERF outperforms other architectures in terms of all metrics, with KAN close second (between 1.1\% and 16\% worse). Whereas, for HyperSound, KAN is better in 4 out of 6 metrics (between 1.3\% and 28.5\%) and worse in two (0.6\% and 11\%). For all the experiments, SIREN performs the worst.

\begin{table}[h]
    \centering
    \caption{Effectiveness of \our{} with different target network architectures.}
    \label{table:our-target-comparison}
    \begin{tabular}{@{}l@{\;}c@{\;}c@{\;}c@{\;}c@{\;}c@{\;}c@{\;}c@{}}
        \toprule
        \textbf{Target} & MSE & PSNR & LSD & SI-SNR & PESQ & STOI & CDPAM \\
        \textbf{Network} & $\rightarrow 0$ & $\rightarrow \infty$ & $\rightarrow 0$ & $\rightarrow 100$ & $\rightarrow 4.5$ & $\rightarrow 1$ & $\rightarrow 0$ \\
        \midrule
        \textit{NERF} & \textbf{0.006} & \textbf{28.07} & \textbf{1.50} & \textbf{4.81} & \textbf{1.38} & \textbf{0.78} & \textbf{0.31} \\
        \textit{KAN} & 0.007 & 27.76 & 1.55 & 4.26 & 1.36 & 0.77 & 0.36 \\
        \textit{SIREN} & 0.025 & 22.02 & 1.74 & -9.40 & 1.18 & 0.64 & 0.31 \\
        \bottomrule
    \end{tabular}
\end{table}

\begin{table}[h]
    \centering
    \caption{Performance of HyperSound with different target network architectures. Results for SIREN and NERF are taken from the original paper, while we train the model using the KAN target network architecture ourselves. A target network with a significantly lower number of parameters (19K) is used in this case, as it achieves better results.}
    \label{table:hs-target-comparison}
    \begin{tabular}{@{}l@{\;\;}c@{\;\;}c@{\;\;}c@{\;\;}c@{\;\;}c@{\;\;}c@{\;\;}c@{}}
        \toprule
        \textbf{Target} & MSE & LSD & SI-SNR & PESQ & STOI & CDPAM \\
        \textbf{Network} & $\rightarrow 0$ & $\rightarrow 0$ & $\rightarrow 100$ & $\rightarrow 4.5$ & $\rightarrow 1$ & $\rightarrow 0$ \\
        \midrule
        \textit{NERF} & 0.009 & \textbf{1.61} & 2.99 & 1.27 & 0.76 & \textbf{0.36} \\
        \textit{KAN} & \textbf{0.007} & 1.62 & \textbf{3.98} & \textbf{1.29} & \textbf{0.77} & 0.40 \\
        \textit{SIREN} & 0.018 & 3.56 & -3.66 & 1.19 & 0.65 & 0.3 \\
        \bottomrule
    \end{tabular}
\end{table}

\paragraph{\textbf{Influence of encoder architecture}}
Table~\ref{table:encoder-comparison} shows the impact of using different encoder architectures, mentioned in Section~\ref{sec:related}. SoundStream, SNAC, and Spectrogram Encoder are trained from scratch. In the case of Whisper, we base on the pre-trained checkpoints - \textit{tiny.en} and \textit{large-v2} and finetune using the LoRA approach \cite{lora}. Among the encoders considered, SoundStream achieves the best results in terms of distance-based metrics such as PSNR, while Spectrogram Encoder excels in perceptual ones. SNAC doesn't yield poor results, but it is essentially a less effective version of SoundStream. The Whisper models, by contrast, show significantly lower PSNR along with weaker PESQ score. 

\begin{table}[!h]
    \centering
    \caption{\our{} results with different encoders.}
    \label{table:encoder-comparison}
    \begin{tabular}{@{}l@{\;\;}c@{\;\;}c@{\;\;}c@{\;\;}c@{\;\;}c@{\;\;}c@{\;\;}c@{}}
        \toprule
        \textbf{Encoder} & MSE & PSNR & LSD & SI-SNR & PESQ & STOI & CDPAM \\
        & $\rightarrow 0$ & $\rightarrow \infty$ & $\rightarrow 0$ & $\rightarrow 100$ & $\rightarrow 4.5$ & $\rightarrow 1$ & $\rightarrow 0$ \\
        \midrule
        \textit{SoundStream} & \textbf{0.006} & \textbf{28.07} & 1.50 & \textbf{4.81} & 1.38 & 0.78 & 0.31 \\
        \textit{Whisper L} & 0.042 & 19.72 & 1.40 & -27.67 & 1.30 & 0.75 & 0.31 \\
        \textit{Whisper T} & 0.039 & 20.10 & 1.40 & -29.08 & 1.25 & 0.72 & 0.33  \\
        \textit{Spectral} & 0.045 & 19.46 & \textbf{1.35} & -21.76 & \textbf{1.49} & \textbf{0.80} & \textbf{0.29} \\
        \textit{SNAC} & 0.007 & 27.55 & 1.56 & 4.03 & 1.30 & 0.77 & 0.32 \\
        \bottomrule
    \end{tabular}
\end{table}

\paragraph{\textbf{Effect of dataset variability}}
We extend the results for our \our{} model by training it on three additional speech datasets - \textit{noisy} VCTK \cite{noisyvctk}, a version of the VCTK dataset with added noise, LibriSpeech \cite{librispeech} and LJ Speech~\cite{ljspeech}. Moreover, it is also evaluated on GTZAN~\cite{gtzan} and UrbanSound8K~\cite{urbansound}, which represent music and environmental noise, respectively. We follow the same setup as in our main experiment, with one exception—in this experiment, we use a hypernetwork MLP with 41.2M parameters instead of the standard one. In case of Librispeech, we train on the \textit{train-clean-100} partition and use the \textit{test-other} partition as the validation set. Since PESQ, STOI and CDPAM are suitable only for clean speech datasets, they are omitted from the \our{} model assessment on \textit{noisy} VCTK, GTZAN and UrbanSound8K. All obtained results are presented in Table \ref{table:fewsound-dataset-comp}. 
The results indicate that \our{} achieves lower MSE and higher SI-SNR on speech datasets than on music or environmental noise, likely due to encoder limitations in handling non-speech signals. Between the KAN and NeRF targets, KAN consistently outperforms in LSD and PESQ across all datasets. For SI-SNR and PSNR, performance is comparable, e.g., NeRF excels on LibriSpeech, while KAN leads on LJ Speech.

\begin{table}[h]
    \centering
    \caption{\our{} performance across different datasets.}
    \label{table:fewsound-dataset-comp}
    \resizebox{0.47\textwidth}{!}{
    \begin{tabular}{lcccccccc}
        \toprule
        Dataset & Target & MSE & PSNR & LSD & SI-SNR & PESQ & STOI & CDPAM \\
         & Net. & $\rightarrow 0$ & $\rightarrow \infty$ & $\rightarrow 0$ & $\rightarrow 100$ & $\rightarrow 4.5$ & $\rightarrow 1$ & $\rightarrow 0$ \\
        \midrule
        \textit{VCTK}
        & \textit{NERF}
        & \textbf{0.006}
        & \textbf{28.39}
        & 1.50 
        & 5.03 
        & 1.41 
        & \textbf{0.79}
        & \textbf{0.30} 
        \\
        & \textit{KAN}
        & \textbf{0.006}
        & 28.35
        & \textbf{1.46}
        & \textbf{5.11}
        & \textbf{1.45}
        & \textbf{0.79}
        & 0.34
        \\
        \midrule
        \textit{noisy}
        & \textit{NERF} 
        & \textbf{0.013}
        & \textbf{24.76}
        & 1.42
        & 2.15
        & -
        & -
        & -
        \\
        \textit{VCTK}
        & \textit{KAN}
        & \textbf{0.013}
        & 24.69
        & \textbf{1.31}
        & \textbf{3.31}
        & -
        & -
        & -
        \\
        \midrule
        \textit{Libri}
        & \textit{NERF} 
        & \textbf{0.016}
        & \textbf{23.8}
        & 2.42
        & \textbf{-2.18}
        & 1.26
        & 0.71
        & 0.34
        \\
        \textit{Speech}
        & \textit{KAN} 
        & 0.018
        & 23.47
        & \textbf{1.93}
        & -3.21
        & \textbf{1.31}
        & \textbf{0.73}
        & \textbf{0.31}
        \\
        \midrule
        \textit{LJ}
        & \textit{NERF} 
        & 0.013
        & 24.80
        & 1.66
        & -0.72
        & 1.22
        & 0.74
        & 0.38
        \\
        \textit{Speech}
        & \textit{KAN}
        & \textbf{0.012}
        & \textbf{25.11}
        & \textbf{1.51}
        & \textbf{-0.43}
        & \textbf{1.24}
        & \textbf{0.76}
        & \textbf{0.35}
        \\
        \midrule
        \textit{GTZAN}
        & \textit{NERF} 
        & 0.031
        & 21.21
        & 1.49
        & -4.23
        & -
        & -
        & -
        \\
        & \textit{KAN}
        & \textbf{0.028}
        & \textbf{21.59}
        & \textbf{1.34}
        & \textbf{-3.50}
        & -
        & -
        & -
        \\
        \midrule
        \textit{Urban}
        & \textit{NERF} 
        & 0.051
        & 18.94
        & 1.28
        & -12.93
        & -
        & -
        & -
        \\
        \textit{Sound-8k}
        & \textit{KAN}
        & \textbf{0.049}
        & \textbf{19.15}
        & \textbf{1.22}
        & \textbf{-11.86}
        & -
        & -
        & -
        \\
        \bottomrule
    \end{tabular}
    }
\end{table}

\subsection{KAN Ablation Study}\label{p:ablation}

\begin{table}[!htbp]
    \footnotesize
    \centering
    \caption{KAN performance on the 1.5~s samples, with respect to the encoding length. The results are averaged across 10 samples.}
    \label{table:endod_len-comparison-1_5}
    \begin{tabular}{@{}l@{\;\;}c@{\;\;}c@{\;\;}c@{\;\;}c@{\;\;}c@{\;\;}c@{\;\;}c@{\;\;}c@{}}
        \toprule
        \textbf{Enc. Len.} & \#Params  & MSE & PSNR & LSD & SI-SNR & PESQ & STOI & CDPAM \\
        & & $\rightarrow 0$ & $\rightarrow \infty$ & $\rightarrow 0$ & $\rightarrow 100$ & $\rightarrow 4.5$ & $\rightarrow 1$ & $\rightarrow 0$ \\
        \toprule
        \textit{2} & 23,016 & 3.3e-3 & 30.72 & 2.39 & 9.03 & 1.50 & 0.76 & 0.37 \\
        \textit{4} & 25,704 & 1.7e-3 & 33.76 & 1.60 & 12.31 & 1.95 & 0.90 & 0.26 \\
        \textit{8} &   31,080  & 7e-4 & 40.10 & \textbf{1.29} & 18.95 & 3.36 & 0.99 & 0.19 \\
        \textit{10} & 33,768 & 5e-4 & 41.12 & 1.39 & 20.01 & 3.49 & 0.99 & 0.16 \\
        \textit{12} & 36,456 & 5e-4 & 42.65 & 1.42 & 21.61 & 3.52 & 0.99 & 0.17 \\
        \textit{16} & 41,832 & \textbf{2e-4} & \textbf{46.86} & 1.42 & \textbf{25.98} & \textbf{3.63} & \textbf{1.00} & \textbf{0.15} \\   
        \textit{24} & 52,584 & 3e-4 & 45.79 & 1.48 & 24.91 & 3.60 & \textbf{1.00} & 0.20 \\
        \bottomrule
    \end{tabular}
\end{table}

\begin{table}[!htbp]
\footnotesize
    \centering
    \caption{KAN results on 1.5~s samples in relation to different number and sizes of hidden layers.}
    \label{table:target_architectures-comparison-1_5}
    \begin{tabular}{@{}l@{\;\;}c@{\;\;}c@{\;\;}c@{\;\;}c@{\;\;}c@{\;\;}c@{\;\;}c@{\;\;}c@{}}
        \toprule
        \textbf{Layer Sizes} & \#Params & MSE & PSNR & LSD & SI-SNR & PESQ & STOI & CDPAM \\
        & & $\rightarrow 0$ & $\rightarrow \infty$ & $\rightarrow 0$ & $\rightarrow 100$ & $\rightarrow 4.5$ & $\rightarrow 1$ & $\rightarrow 0$ \\
        \midrule
               \textit{[24, 12, 6]} & 10,500  & 1.4e-3 & 35.92 & 1.48 & 14.53 & 2.13 & 0.94 & 0.30 \\
               \textit{[24, 24, 12, 6]} &   18,564 & 1.2e-3 & 37.68 & 1.36 & 16.42 & 2.70 & 0.98 & 0.27 \\
               \textit{[48, 12, 6]} &   19,908  & 1e-3 & 38.57 & 1.36 & 17.31 & 2.86 & 0.98 & 0.27 \\
               \textit{[48, 12, 12, 6]} & 21,924  & 1e-3 & 38.29 & 1.36 & 17.08 & 2.96 & 0.98 & 0.26 \\
               \textit{[48, 16, 8]} &  23,408  & 9e-4 & 38.73 & 1.33 & 17.51 & 3.05 & 0.98 & 0.25 \\
               \textit{[48, 24, 12]} &   31,080  & 7e-4 & 40.10 & 1.29 & 18.95 & 3.36 & 0.99 & 0.19 \\
               \textit{[96, 48, 24]} &  102,480  & \textbf{3e-4} & \textbf{43.64} & \textbf{1.17} & \textbf{22.57} & \textbf{4.02} & \textbf{1.00} & \textbf{0.11} \\
        \bottomrule
    \end{tabular}
\end{table}

Next, we analyze the influence of various parameters' values on the KAN's performance. In this ablation study, experiments are also conducted on audio samples of 1.5 seconds and 3 seconds, with each subset limited to only 10 recordings. During our study we keep the following base parameters' values, changing only the analyzed components: \textit{encoding length}~=~8, \textit{grid size}~=~10, \textit{spline order}~=~2, \textit{layer sizes}~=~[48, 24, 12], 
\textit{learning rate}~=~5e-3 and \textit{enable scale spline}~=~True, which controls whether we include the learnable spline weigths matrix $w_s$. Theoretically, these weigths are redundant, as they could be absorbed into the spline itself. In this section, we focus on the 1.5s samples and the most important parameters. The remaining ablation studies can be found in section \ref{sect:ablation-cd}.

Table~\ref{table:endod_len-comparison-1_5} illustrates the impact of the length of the encoding (size of the positional encoding), one of the parameters directly influencing the size of the network. Values smaller than the baseline of 8 result in progressively worse performance; meanwhile, increasing the length provides improvement up to value of 16, where the results are the best for all but one metric (LSD). 

Table~\ref{table:target_architectures-comparison-1_5} shows that increasing hidden layer capacity consistently improves performance, with the [96, 48, 24] configuration performing best. Smaller architectures yield progressively worse results.

Furthermore, Table~\ref{table:spline_order-comparison-1_5} presents a comparison of the varying spline order values (the degree of the applied spline). Increasing the spline order allows creation of more complex spline functions; however, this comes at the cost of additional parameters. In our study, an order of 3 provides the best overall performance across both quantitative (PSNR) and perceptual (PESQ) metrics. Using smaller or larger spline orders leads to a decline in metric scores.

Finally, Table~\ref{table:grid_size-comparison-1_5} provides insights into the influence of grid size (number of
knots defining the spline) on KAN's learning process. Similarly to encoding length and spline order, an increase in grid size leads to a higher number of INR parameters. While a larger spline order eventually results in a decline in reconstruction quality, this effect is not observed with grid size. Extra parameters enhance the expressiveness of the INR, improving all metrics except LSD.

\begin{table}[!h]
    \centering
    \caption{KAN spline order~(O) for 1.5~s audio recordings}
    \label{table:spline_order-comparison-1_5}
    \begin{tabular}{@{}l@{\;\;}c@{\;\;}c@{\;\;}c@{\;\;}c@{\;\;}c@{\;\;}c@{\;\;}c@{\;\;}c@{}}
        \toprule
        \textbf{O} & Params  & MSE & PSNR & LSD & SI-SNR & PESQ & STOI & CDPAM \\
        & & $\rightarrow 0$ & $\rightarrow \infty$ & $\rightarrow 0$ & $\rightarrow 100$ & $\rightarrow 4.5$ & $\rightarrow 1$ & $\rightarrow 0$ \\
        \midrule
               \textit{1} &   28,860 & 9e-4 & 38.83 & 1.53 & 17.59 & 3.01 & 0.98 & 0.27 \\
               \textit{2} &   31,080 & \textbf{7e-4} & 40.10 & 1.29 & 18.95 & 3.36 & \textbf{0.99} & 0.19 \\
               \textit{3} &   33,300 & \textbf{7e-4} & \textbf{40.43} & 1.23 & \textbf{19.28} & \textbf{3.46} & \textbf{0.99} & \textbf{0.18} \\
               \textit{5} &   37,740 & 1e-3 & 38.87 & \textbf{1.22} & 17.66 & 3.38 & \textbf{0.99} & 0.19 \\
               \textit{7} &   42,180 & 9e-4 & 38.91 & 1.23 & 17.71 & 3.21 & 0.98 & 0.21 \\
        \bottomrule
    \end{tabular}
\end{table}

\begin{table}[!h]
    \centering
    \caption{Impact of spline function grid size~(G) on 1.5~s reconstructions}
    \label{table:grid_size-comparison-1_5}
    \begin{tabular}{@{}l@{\;\;}c@{\;\;}c@{\;\;}c@{\;\;}c@{\;\;}c@{\;\;}c@{\;\;}c@{\;\;}c@{}}
        \toprule
         \textbf{G} & Params & MSE & PSNR & LSD & SI-SNR & PESQ & STOI & CDPAM \\
        & & $\rightarrow 0$ & $\rightarrow \infty$ & $\rightarrow 0$ & $\rightarrow 100$ & $\rightarrow 4.5$ & $\rightarrow 1$ & $\rightarrow 0$ \\
        \midrule
               \textit{1} &   11,100 & 3e-3 & 30.65 & 2.00 &  8.99 & 1.65 & 0.80 & 0.43 \\
               \textit{5} &   19,980 & 1e-3 & 37.33 & 1.36 & 16.01 & 2.61 & 0.97 & 0.27 \\
               \textit{10} &  31,080 & 7e-4 & 40.10 & \textbf{1.29} & 18.95 & 3.36 & 0.99 & 0.19 \\
               \textit{12} &   35,520 & 5e-4 & 42.34 & 1.30 & 21.24 & 3.65 & 0.99 & 0.16 \\
               \textit{17} &   46,620 & \textbf{3e-4} & \textbf{44.78} & 1.31 & \textbf{23.78} & \textbf{3.81} & \textbf{1.00} & \textbf{0.13} \\
        \bottomrule
    \end{tabular}
\end{table}

\subsection{Frequency-based analysis}

\paragraph{\textbf{Spectral Analysis}}
To demonstrate KAN's improvements over previous INRs in high-frequency information of the reconstruction, we designed an additional experiment. We can treat the absolute values of the FFT of the audio signal as probability distribution of the base frequencies. Then, we can compare distributions of the ground truth and the reconstruction using Wasserstein distance. By using $[0, 1, 4, ... (len(audio) - 1)^2]$ as the support of the sample instead of intuitive $[0, 1, 2, ... len(audio) - 1]$, we can make the metric more biased towards high frequencies. Using this metric in our standard experimental setting for INRs, we obtain the following results, presented in Table \ref{table:wasserstein}. They indicate that KAN achieves the lowest score of 79.72 among all baseline methods. The second-best outcome is obtained by WIRE, while RFF performs the worst.

\begin{table}
    \caption{Wasserstein Distance (WD) of reconstruction and ground truth frequency distributions with the high-frequency biased support. For readability purposes the actual values are scaled down by a constant factor of 32768, that is the length of the sample (lower score is better).}
    \label{table:wasserstein}
    \centering
    \begin{tabular}{ccccccc}
        \toprule
         & \textit{NERF} & \textit{SIREN} & \textit{FINER} & \textit{WIRE} & \textit{RFF} & \textit{KAN} \\
         \midrule
         WD & 184.79 & 218.54 & 205.75 & 131.56 & 362.98 & \textbf{79.72} \\
         \bottomrule
    \end{tabular}
\end{table}

\begin{figure*}[!h]
    \centering
    \includegraphics[width=2.1\columnwidth]{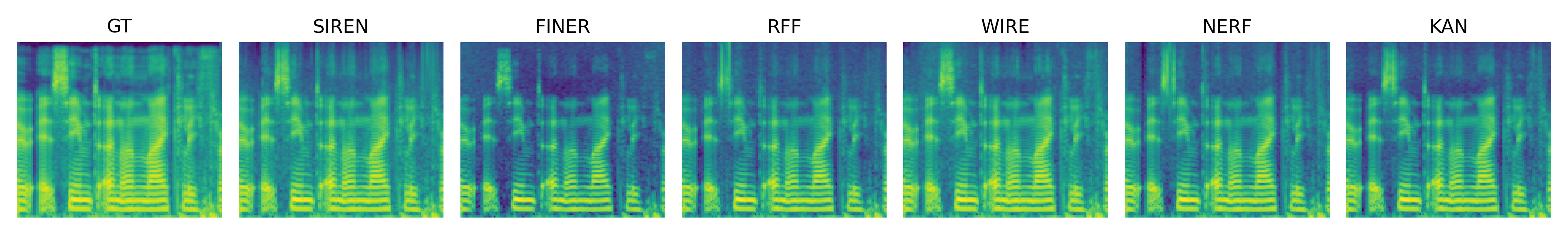}
    \Description{The comparison of spectrograms. We compare the original sample (GT) with the waveforms reconstructed using different INR approaches.}
    \caption{The comparison of spectrograms. We compare the original sample (GT) with the waveforms reconstructed using different INR approaches.}
    \label{fig:spectrograms}
\end{figure*}

\paragraph{\textbf{Spectrograms}}

The spectrograms shown in Fig.~\ref{fig:spectrograms} illustrate differences in the fidelity of sound reconstruction achieved by various INRs compared to the ground truth. Across all methods, the harmonic structures in the low-frequency range are generally well-preserved, reflecting their ability to capture fundamental features of the audio signal. However, significant variations emerge in the representation of high-frequency components, where most INRs lower the intensity of the components. It is particularly visible with FINER. Moreover, all of the INRs (with KAN and NERF to the least degree) demonstrate a tendency to over-smooth the higher intensities in high-frequency components, which results in adding white noise. Finally, one can note the lower average intensity in the reconstructed audio.

\section{Conclusion}
This work demonstrates the effectiveness of Kolmogorov-Arnold Networks (KANs) as implicit neural representations for audio signals, offering high reconstruction accuracy when integrated with hypernetwork-based architectures. By evaluating KANs against classical INR models like NERF, FINER, WIRE, RFF, and SIREN, we highlight their potential as a foundational approach for efficient audio modeling, resulting in top-2 mean in 5 out of 7 considered quality metrics for 1.5~s audio, and top-2 in all 7 for 3~s audio.
Additionally, we introduce a new hypernetwork framework called FewSound, and show that by integrating KAN performance improves significantly, surpassing HyperSound with a 33.3\% reduction in MSE, a 60.87\% increase in SI-SNR, and an 8.66\% in PESQ. The results suggest that KANs can serve as a powerful tool for audio representation, enabling adaptive processing across various datasets and tasks.

As the experiments used a fixed 22{,}050~Hz sampling rate, exploring alternative sampling strategies remains a future work. Integration with modern neural codecs and speech encoders (Tab.~\ref{table:encoder-comparison}) could be extended, particularly for multilingual scenarios. Moreover, our evaluation focused on English due to dataset availability and lacked complex auditory conditions (e.g., multi-speaker, emotional speech, diverse scenes), suggesting avenues for future research.  It is important to note that while KANs demonstrated efficiency in certain tasks, their computational demands may grow significantly for longer sequences or large-scale applications.

Future research should extend KAN-based models to multilingual datasets and varied linguistic contexts to improve generalization. Analyzing the effect of sample length may offer insight into balancing fidelity and computational efficiency. Architectural optimizations could reduce resource usage while preserving accuracy, enhancing scalability. Integrating KANs with multimodal frameworks, e.g., facial expressions or gestures could enable advanced audio-visual synthesis, such as avatar generation, especially given the use of hypernetworks in visual domains~\cite{PRZEWIEZLIKOWSKI2024128179,kania2024freshfrequencyshiftingaccelerated}. Developing interpretable models would improve transparency and trust, supporting real-world deployment and legal compliance. Lastly, enhancing compatibility with diverse audio encoding systems could boost reconstruction quality and expand KANs’ practical utility.

\begin{acks}
This work was funded by several institutions. P. Marszałek was supported by the National Center of Science (Poland) Grant No. \\ 2023/50/E/ST6/00169. P. Kawa and P. Syga received support from the Department of Artificial Intelligence, Wrocław University of Science and Technology. P. Spurek and M. Rut were supported by the National Center of Science (Poland) Grant No. 2023/50/E/ST6/00068.
\end{acks}

\appendix

\section{Additional experiments}

\subsection{Extended ablation study}\label{sect:ablation-cd}
At the beginning of this section we conclude the ablation study (presented initially in Sect. ~\ref{p:ablation}). We show how KAN performance on 3~s audio samples is influenced by \textit{encoding length}, \textit{layer sizes}, \textit{spline order} and ~\textit{spline function grid size} parameters. We consider the same hyperparameters as used for the 1.5~s experiments.

Table~\ref{table:endod_len-comparison-3} shows the impact of different \textit{encoding lengths} on 3~s utterances. The absolute performance is generally lower across all metrics in relation to 1.5~s samples, e.g. the best PESQ decreasing from 3.62 to 2.90 and the best SI-SNR decreasing from 25.97 to 17.17. The results are much less conclusive than in case of 1.5~s, where length 16 was clearly optimal across most metrics. However, the majority of metrics (MSE, PSNR, SI-SNR and PESQ, STOI) are the best for encoding lengths of 24 and 16, respectively.

\begin{table}[!htbp]
\footnotesize
    \centering
    \caption{KAN performance on the 3~s samples, with respect to the encoding length.}
    \label{table:endod_len-comparison-3}
    \begin{tabular}{@{}l@{\;\;}c@{\;\;}c@{\;\;}c@{\;\;}c@{\;\;}c@{\;\;}c@{\;\;}c@{\;\;}c@{}}
        \toprule
        \textbf{Encoding} & \#Params &  MSE & PSNR & LSD & SI-SNR & PESQ & STOI & CDPAM \\
        \textbf{Length}& & $\rightarrow 0$ & $\rightarrow \infty$ & $\rightarrow 0$ & $\rightarrow 100$ & $\rightarrow 4.5$ & $\rightarrow 1$ & $\rightarrow 0$ \\
        \toprule
        \textit{2} & 23,016 & 3.6e-3 & 30.46 & 3.31 & 4.26 & 1.21 & 0.66 & 0.42 \\
        \textit{4} & 25,704 & 1.1e-3 & 35.41 & 2.03 & 10.31 & 1.55 & 0.82 & \textbf{0.29} \\
        \textit{8} & 31,080 & 6e-4 & 38.93 & \textbf{1.27} & 14.03 & 2.50 & 0.97 & 0.36 \\
        \textit{10} & 33,768 & 6e-4 & 39.40 & 1.32 & 14.55 & 2.83 & 0.97 & 0.38 \\
        \textit{12} & 36,456 & 5e-4 & 39.66 & 1.37 & 14.79 & 2.78 & \textbf{0.98} & 0.37 \\
        \textit{16} & 41,832 & 4e-4 & 40.78 & 1.44 & 15.95 & \textbf{2.91} & \textbf{0.98} & 0.35 \\
        \textit{24} & 52,584 & \textbf{3e-4} & \textbf{41.93} & 1.60 & \textbf{17.18} & 2.84 & \textbf{0.98} & 0.34 \\
        \bottomrule
    \end{tabular}
\end{table}

\begin{table}[!htbp]
\footnotesize
    \centering
    \caption{KAN results on 3~s samples in relation to different number and sizes of hidden layers.}
    \label{table:target_architectures-comparison-3}
    \begin{tabular}{@{}l@{\;\;}c@{\;\;}c@{\;\;}c@{\;\;}c@{\;\;}c@{\;\;}c@{\;\;}c@{\;\;}c@{}}
        \toprule
        \textbf{Layer Sizes} & \#Params & MSE & PSNR & LSD & SI-SNR & PESQ & STOI & CDPAM \\
        & & $\rightarrow 0$ & $\rightarrow \infty$ & $\rightarrow 0$ & $\rightarrow 100$ & $\rightarrow 4.5$ & $\rightarrow 1$ & $\rightarrow 0$ \\
        \midrule
       \textit{[24, 12, 6]}     & 10,500  & 7e-4 & 37.57 & 1.49 & 12.66 & 1.72 & 0.89 & 0.32 \\
       \textit{[24, 24, 12, 6]} & 18,564  & 6e-4 & 38.42 & 1.38 & 13.52 & 1.91 & 0.94 & 0.38 \\
       \textit{[48, 12, 6]}     & 19,908  & 8e-4 & 38.01 & 1.33 & 13.03 & 2.16 & 0.94 & 0.37 \\
       \textit{[48, 12, 12, 6]} & 21,924  & 6e-4 & 38.31 & 1.34 & 13.40 & 2.18 & 0.95 & 0.39 \\
       \textit{[48, 16, 8]}     & 23,408  & 7e-4 & 38.29 & 1.31 & 13.35 & 2.30 & 0.96 & 0.38 \\
       \textit{[48, 24, 12]}    & 31,080  & 6e-4  & 38.93 & 1.27 & 14.03 & 2.50 & 0.97 & 0.36 \\
       \textit{[96, 48, 24]}    & 102,480 & \textbf{5e-4} & \textbf{40.96} & \textbf{1.18} & \textbf{16.12} & \textbf{3.52} & \textbf{0.99} & \textbf{0.16} \\
        \bottomrule
    \end{tabular}
\end{table}

Table~\ref{table:target_architectures-comparison-3} shows how sizes and depths of hidden layers influence results on 3~s samples. Similarly to 1.5~s (Table~\ref{table:target_architectures-comparison-1_5}) the best performance is achieved by the largest network composed of [96, 48, 24] layers. The performance, however, does not improve with the increasing size as gradually as for 1.5~s samples.

Compared to the 1.5~s case, where a \textit{spline order} of 3 is clearly optimal, its advantage is less evident in Tab.~\ref{table:spline_order-comparison-3}. KAN with a spline order of 5 attains higher PSNR and SI-SNR, while remaining competitive in PESQ and STOI.

\begin{table}[!htbp]
\footnotesize
    \centering
    \caption{KAN effectiveness in terms of diverse spline order sizes and 3~s long input}
    \label{table:spline_order-comparison-3}
    \begin{tabular}{@{}l@{\;\;}c@{\;\;}c@{\;\;}c@{\;\;}c@{\;\;}c@{\;\;}c@{\;\;}c@{\;\;}c@{\;\;}c@{}}
        \toprule
        \textbf{Spline} & \#Params & MSE & PSNR & LSD & SI-SNR & PESQ & STOI & CDPAM \\
        \textbf{Order} & & $\rightarrow 0$ & $\rightarrow \infty$ & $\rightarrow 0$ & $\rightarrow 100$ & $\rightarrow 4.5$ & $\rightarrow 1$ & $\rightarrow 0$ \\
        \midrule
              \textit{1} & 28,860 & \textbf{6e-4} & 38.50 & 1.47 & 13.61 & 2.29 & 0.95 & \textbf{0.33} \\
              \textit{2} & 31,080 & \textbf{6e-4} & 38.93 & 1.27 & 14.03 & 2.50 & \textbf{0.97} & 0.36 \\
              \textit{3} & 33,300 & 8e-4 & 38.66 & \textbf{1.25} & 13.68 & \textbf{2.56} & \textbf{0.97} & 0.36 \\
              \textit{5} & 37,740 & \textbf{6e-4} & \textbf{39.09} & 1.27 & \textbf{14.19} & 2.45 & \textbf{0.97} & 0.34 \\
              \textit{7} & 42,180 & \textbf{6e-4} & 38.88 & 1.32 & 13.97 & 2.20 & 0.96 & 0.34 \\
        \bottomrule
    \end{tabular}
\end{table}


The grid size results (cf.~Tab.~\ref{table:grid_size-comparison-3}) show trends similar to the 1.5~s case (Tab.~\ref{table:grid_size-comparison-1_5}), increasing grid size improves all metrics except LSD.

\begin{table}[!htbp]
\footnotesize
    \centering
    \caption{KAN effectiveness in terms of diverse grid size and 3~s long input}
    \label{table:grid_size-comparison-3}
    \begin{tabular}{@{}l@{\;\;}c@{\;\;}c@{\;\;}c@{\;\;}c@{\;\;}c@{\;\;}c@{\;\;}c@{\;\;}c@{\;\;}c@{}}
        \toprule
        \textbf{Grid size} & \#Params & MSE & PSNR & LSD & SI-SNR & PESQ & STOI & CDPAM \\
       & & $\rightarrow 0$ & $\rightarrow \infty$ & $\rightarrow 0$ & $\rightarrow 100$ & $\rightarrow 4.5$ & $\rightarrow 1$ & $\rightarrow 0$ \\
        \midrule
               1 & 11,100 & 2.9e-3 & 31.34 & 2.00 &  5.46 & 1.38 & 0.77 & 0.40 \\
               5 & 19,980 & 7e-4 & 38.11 & 1.47 & 13.21 & 1.97 & 0.92 & \textbf{0.24} \\
               10 & 31,080 & 6e-4 & 38.93 & \textbf{1.27} & 14.03 & 2.50 & \textbf{0.97} & 0.36 \\
               12 & 35,520 & \textbf{5e-4} & 39.49 & \textbf{1.27} & 14.64 & 2.67 & \textbf{0.97} & 0.32 \\
               17 & \textbf{46,620} & 6e-4 & \textbf{39.73} & 1.32 & \textbf{14.81} & \textbf{3.07} & \textbf{0.97} & 0.27 \\
        \bottomrule
    \end{tabular}
\end{table}

At the end of this paragraph, we include ablation studies for the remaining hyperparameters. Table~\ref{table:learning_rate-comparison-merged} shows the comparison of \textit{learning rates}. A value of 0.001 yields the worst scores across all entries in the table, with the exception of the CDPAM metric. A learning rate of 0.01 achieves the best results in the PSNR and SI-SNR metrics, which are quantitative, while a learning rate near 0.006 produces the highest perceptual scores. Generally, KAN performs well and stably with higher learning rate setting. It can be considered its advantage over networks like SIREN, which are sensitive to the choice of the learning rate and their learning process can collapse when too high values are used.

\begin{table}[htbp]
\footnotesize
\centering
\caption{Effect of various learning rate values on 1.5~s and 3~s recordings.}
\label{table:learning_rate-comparison-merged}
\begin{tabular}{@{}l@{\;\;}c@{\;\;}c@{\;\;}c@{\;\;}c@{\;\;}c@{\;\;}c@{\;\;}c@{\;\;}c@{\;\;}c@{}}
    \toprule
    \textbf{Learning} & \#Params & MSE & PSNR & LSD & SI-SNR & PESQ & STOI & CDPAM \\
    \textbf{Rate} & & $\rightarrow 0$ & $\rightarrow \infty$ & $\rightarrow 0$ & $\rightarrow 100$ & $\rightarrow 4.5$ & $\rightarrow 1$ & $\rightarrow 0$ \\
    \midrule
    \multicolumn{9}{c}{\textbf{Sample Length: 1.5 s}} \\
    \midrule
    \textit{0.01}  & 31,080 & \textbf{7e-4} & \textbf{40.56} & 1.37 & \textbf{19.47} & 3.28 & \textbf{0.99} & 0.19 \\
    \textit{0.007} & 31,080 & \textbf{7e-4} & 40.52 & 1.33 & 19.36 & \textbf{3.39} & \textbf{0.99} & 0.19 \\
    \textit{0.006} & 31,080 & \textbf{7e-4} & 40.02 & 1.31 & 18.92 & 3.35 & \textbf{0.99} & \textbf{0.18} \\
    \textit{0.005} & 31,080 & \textbf{7e-4} & 40.10 & \textbf{1.29} & 18.95 & 3.36 & \textbf{0.99} & 0.19 \\
    \textit{0.001} & 31,080 & 1.1e-3 & 37.89 & 1.36 & 16.63 & 2.60 & 0.97 & 0.23 \\
    \midrule
    \multicolumn{9}{c}{\textbf{Sample Length: 3 s}} \\
    \midrule
    0.01  & 31,080 & \textbf{6e-4} & \textbf{39.50} & 1.32 & \textbf{14.64} & 2.56 & \textbf{0.97} & 0.39 \\
    0.007 & 31,080 & \textbf{6e-4} & 39.23 & 1.28 & 14.34 & \textbf{2.57} & \textbf{0.97} & 0.37 \\
    0.006 & 31,080 & \textbf{6e-4} & 39.10 & 1.28 & 14.22 & 2.55 & \textbf{0.97} & 0.36 \\
    0.005 & 31,080 & \textbf{6e-4} & 38.93 & \textbf{1.27} & 14.03 & 2.50 & \textbf{0.97} & 0.36 \\
    0.001 & 31,080 & 7e-4 & 37.76 & 1.47 & 12.80 & 1.93 & 0.93 & \textbf{0.30} \\
    \bottomrule
\end{tabular}
\end{table}

Table~\ref{table:enable_scale-comparison-merged} shows the comparison of performance of KAN depending on whether the \textit{enable scale spline} flag is set to False or True. The former results in skipping the theoretically redundant weights. However, it leads to a drop in all metric scores except MSE, with perceptual metrics being affected the most.

\begin{table}[htbp]
\footnotesize
\centering
\caption{Effect of enabling the scale spline flag for 1.5~s and 3~s recordings.}
\label{table:enable_scale-comparison-merged}
\begin{tabular}{@{}l@{\;\;}c@{\;\;}c@{\;\;}c@{\;\;}c@{\;\;}c@{\;\;}c@{\;\;}c@{\;\;}c@{}}
    \toprule
    \textbf{Scale} & \#Params & MSE & PSNR & LSD & SI-SNR & PESQ & STOI & CDPAM \\
    \textbf{Spline} & & $\rightarrow 0$ & $\rightarrow \infty$ & $\rightarrow 0$ & $\rightarrow 100$ & $\rightarrow 4.5$ & $\rightarrow 1$ & $\rightarrow 0$ \\
    \midrule
    \multicolumn{9}{c}{\textbf{Sample Length: 1.5 s}} \\
    \midrule
    \textit{False} & 28,860 & \textbf{7e-4} & 39.41 & 1.53 & 18.23 & 2.74 & 0.97 & 0.28 \\
    \textit{True}  & 31,080 & \textbf{7e-4} & \textbf{40.10} & \textbf{1.29} & \textbf{18.95} & \textbf{3.36} & \textbf{0.99} & \textbf{0.19} \\
    \midrule
    \multicolumn{9}{c}{\textbf{Sample Length: 3 s}} \\
    \midrule
    \textit{False} & 28,860 & 7e-4 & 38.10 & 1.48 & 13.23 & 2.20 & 0.94 & 0.42 \\
    \textit{True}  & 31,080 & \textbf{6e-4} & \textbf{38.93} & \textbf{1.27} & \textbf{14.03} & \textbf{2.50} & \textbf{0.97} & \textbf{0.36} \\
    \bottomrule
\end{tabular}
\end{table}

\subsection{INR}
\label{sec:additional_inr}
To show that KAN performs well across diverse types of audio, we train INRs on samples from the two aforementioned datasets - GTZAN~\cite{gtzan} and UrbanSound8K~\cite{urbansound}.
We chose 100 samples for each dataset, 10 samples per genre in case of GTZAN and 10 samples per split in case of UrbanSound8K. The recordings were pre-processed in an analogous way to our main experiment - resampled to 22.05 kHz and cropped to 32,768 samples, resulting in audio of length 1.5~s. The results are presented in Tables \ref{table:inr-comparison-gtzan} and \ref{table:inr-comparison-urbansound}. While FINER achieves the best PSNR and SI-SNR, it is the worst with respect to LSD. WIRE shows the best overall performance when all metrics are taken into account, followed by SIREN and KAN, both achieving similar results.

\begin{table}
\footnotesize
    \centering
    \caption{Reconstruction metrics for GTZAN dataset. We report mean and standard deviation across 100 samples, 10 samples per genre.}
    \label{table:inr-comparison-gtzan}
    \begin{tabular}{lcccc}
        \toprule
        \textbf{ INR (\# params)} & MSE & PSNR & LSD & SI-SNR \\
        & $\rightarrow 0$ & $\rightarrow \infty$ & $\rightarrow 0$ & $\rightarrow 100$ \\
        \midrule
        \textit{NERF (33,261)}
        & 3.6e-3 $\pm$ 4.5e-3
        & 32.70 $\pm$ 5.12 
        & 1.21 $\pm$ 0.42 
        & 13.16 $\pm$ 5.05 
        \\
        \textit{SIREN (31,191)} 
        & 3.7e-3 $\pm$ 2.5e-2
        & 37.13 $\pm$ 5.18
        & 1.10 $\pm$ 0.38
        & 17.05 $\pm$ 8.37
        \\
        \textit{FINER (31,191)} 
        & 1.0e-3 $\pm$ 2.5e-3
        & 40.89 $\pm$ 5.60
        & 1.23 $\pm$ 0.53
        & 21.36 $\pm$ 4.86
        \\
        \textit{WIRE (31,191)} 
        & 2.0e-3 $\pm$ 6.2e-3
        & 37.13 $\pm$ 5.87
        & 1.03  $\pm$ 0.38
        & 17.50 $\pm$ 5.59
        \\
        \textit{RFF (39,201)} 
        & 4.9e-3 $\pm$ 2.1e-2
        & 32.94 $\pm$ 4.86
        & 1.25 $\pm$ 0.42
        & 13.63 $\pm$ 4.89
        \\
        \textit{\textbf{KAN} (33,768)}
        & 1.6e-3 $\pm$ 2.8e-3
        & 36.98 $\pm$ 5.78 
        & 1.14 $\pm$ 0.45 
        & 17.46 $\pm$ 5.87 
        \\
        \bottomrule
        \end{tabular}
\end{table}

\begin{table}
\footnotesize
    \centering
    \caption{Reconstruction metrics for UrbanSound8K. We report mean and standard deviation across 100 samples, 10 samples per split.}
    \label{table:inr-comparison-urbansound}
    \resizebox{0.47\textwidth}{!}{
        \begin{tabular}{lcccc}
            \toprule
            \textbf{ INR (\# params)} & MSE & PSNR & LSD & SI-SNR \\
            & $\rightarrow 0$ & $\rightarrow \infty$ & $\rightarrow 0$ & $\rightarrow 100$ \\
            \midrule
            \textit{NERF (33,261)}
            & 3e-3 $\pm$ 4.9e-3
            & 34.23 $\pm$ 5.35 
            & 1.25 $\pm$ 0.52 
            & 14.81 $\pm$ 6.33 
            \\
            \textit{SIREN (31,191)}
            & 1.8e-3 $\pm$ 4.8e-3
            & 37.77 $\pm$ 5.79 
            & 1.14 $\pm$ 0.46 
            & 17.15 $\pm$ 12.58 
            \\
            \textit{FINER (31,191)}
            & 7.1e-4 $\pm$ 1.3e-3
            & 40.46 $\pm$ 5.07 
            & 1.27 $\pm$ 0.52 
            & 20.98 $\pm$ 4.23 
            \\
            \textit{WIRE (31,191)}
            & 1.6e-3 $\pm$ 2.7e-3
            & 38.28 $\pm$ 6.74 
            & 1.11 $\pm$ 0.45 
            & 18.65 $\pm$ 7.49 
            \\
            \textit{RFF (39,201)}
            & 4.1e-3 $\pm$ 7.2e-3
            & 33.47 $\pm$ 6.05 
            & 1.23 $\pm$ 0.51 
            & 14.09 $\pm$ 7.41 
            \\
            \textit{\textbf{KAN} (33,768)}
            & 1.6e-3 $\pm$ 2.1e-3
            & 36.92 $\pm$ 6.02 
            & 1.15 $\pm$ 0.49 
            & 17.37 $\pm$ 6.15 
            \\
            \bottomrule
        \end{tabular}
    }
\end{table}

\section*{GenAI Usage Disclosure}
The authors declare that generative AI tools were not used in the preparation of the manuscript or the development of the corresponding source code, apart from software tools for the evaluation of spelling and grammatical correctness.

\bibliographystyle{ACM-Reference-Format}
\bibliography{main}

\end{document}